\documentclass[manuscript]{aastex63}
\usepackage{amsmath}
 \usepackage{xcolor}

\received{July 26, 2019} % January 10, 201
\accepted{October 30, 2019}
\published{December 18, 2019}
\submitjournal{ApJ}

\shorttitle{Synchrotron spectrum}
\shortauthors{Derishev \& Aharonian}
\begin{document}

\title{Exact analytical expression for the synchrotron radiation spectrum in the
Gaussian turbulent magnetic field}

\correspondingauthor{Evgeny Derishev}
%\email{xxxx@aas.org}

%\author[xxxx-xxxx-xxxx-xxxx]{Evgeny Derishev} % xxxx-xxxx-xxxx-xxxx  is  ORCID ID
\author[0000-0002-6761-5515]{Evgeny Derishev}
\affiliation{Institute of Applied Physics RAS, 46 Ulyanov st., 603950 Nizhny Novgorod, Russia}

\author{Felix Aharonian}
\affiliation{Dublin Institute for Advanced Studies, 31 Fitzwilliam Place, Dublin 2, Ireland}
\affiliation{Max-Planck-Institut f\"ur Kernphysik, Saupfercheckweg 1, D-69117 Heidelberg, Germany}

%\collaboration{1}{(AAS Journals Data Scientists collaboration)}



\begin{abstract}

We  demonstrate that the exact solution for the  spectrum of synchrotron radiation from an isotropic population of mono-energetic electrons in turbulent magnetic field with Gaussian distribution of local field strengths can be expressed in the simple analytic  form: \, $\left( \frac{{\rm d} \dot{N}}{{\rm d} \omega} \right)_t
    =  \frac{\alpha}{3} \frac{1}{\gamma^2} \left( 1  + \frac{1}{x^{2/3}} \right)  \exp \left( - 2 x^{2/3} \right)$, where 
$x = \frac{\omega}{\omega_0}\, ; 
\omega_0 = \frac{4}{3} \gamma^2 \frac{eB_0}{m_e c}\, .$
We use this expression to find approximate synchrotron spectra for power-law electron distributions with $\propto \exp\left( -\left[ \gamma/\gamma_0 \right]^\beta\right)$ type high-energy cut-off; the resulting synchrotron spectrum has the exponential cut-off factor with frequency raised to $2\beta/(3\beta+4)$ power in the exponent.
For the power-law electron distribution without high-energy cut-off, we find the coefficient $a_m$ as a function of the power-law index, which results in exact expression for the synchrotron spectrum when using monochromatic (i.e., each electron radiates at frequency $\omega_m = a_m \gamma^2 \, \frac{e B_0}{m_e c}$) approximation. 

\end{abstract}


\section{Introduction} 
\label{sec:intro}

Synchrotron radiation is the most common non-thermal emission mechanism in astrophysics. 
Calculation of its  spectrum involves several steps. One starts with the
expression for synchrotron spectrum (the power emitted per unit frequency) of an individual relativistic electron, moving perpendicular to the field lines of uniform magnetic field. This expression can be found in many textbooks (see, e.g.  Eq.~74.17 in \cite{LandauLifshitz}):
\begin{equation} \label{standard_spectrum}
    \left(\frac{{\rm d} L}{{\rm d} \omega} \right)_u = 
    \frac{\sqrt{3}}{2\pi}\, \frac{e^3 B}{m_e c^2} \,
    F\!\left( \frac{\omega}{\omega_c} \right)\, , \qquad 
    \omega_c = \frac{3}{2} \gamma^2 \frac{eB}{m_e c} \equiv \frac{3}{2} \gamma^2 \omega_\mathrm{_B}  \, .
\end{equation}
Here 
\begin{equation}
    F(x) = x \int_x^\infty K_{5/3}(\xi) \, {\rm d} \xi \, , 
\end{equation}
and $K_{5/3}(\xi)$ is modified Bessel function of the second kind; the subscript $u$ denotes uniform magnetic field. 
Here and below we use $\omega_\mathrm{_B} = \frac{eB}{m_e c}$ to simplify notation.

For a single electron in a turbulent magnetic field,
Eq.~(\ref{standard_spectrum}) should be averaged 
over (1) the pitch angles and (2) the distribution of the local strength of the 
magnetic field. The result is to be convoluted with the electron distribution function. For an isotropic distribution of radiating particles,
the procedure (1) allows exact analytic  expressions in terms of special functions. In particular,  \cite{AngleAveraging} have derived the spectrum expressed in terms of Whittaker functions, while \cite{elementary_approximation} suggested
an alternative formula in terms of modified Bessel functions. Although these solutions are  presented in compact and elegant forms, for practical purposes it is convenient to avoid special functions, i.e., to have analytic approximations containing only elementary functions. The simplest approximation is based on the assumption that electrons emit monochromatic synchrotron photons, whose frequency depends on electrons' energy and the magnetic field strength. This produces reasonably good approximation for featureless, e.g. power-law, electron distributions, but is known to yield wrong results for distributions with high-energy cut-off (e.g., \cite{SyCutoffShape}). Instead, \cite{approxAngleAvg} and \cite{elementary_approximation} have offered simple analytic approximations which deviate from the exact solution less than 3~\% and 0.2~\%, respectively.

Previously published calculations of synchrotron spectrum in turbulent magnetic field dealt with various options for the distribution of local magnetic field strengths, resulting in analytical asymptotic formulas (e.g., \cite{EilekArendt1996, GaussFieldDistribution}). 
For exponential distribution of local field strengths a complex exact expression was derived \citep{ExpFieldDistribution}, for which, however, a simple approximation was proposed. The case of turbulent magnetic field with Gaussian distribution was previously studied numerically \citep{GaussFieldDistribution}.

In this paper we report the finding that both preliminary integrations (1) and (2) can be done analytically for the turbulent magnetic field with Gaussian distribution of the local field strength and isotropic distribution of electrons over pitch angles. Moreover, the final exact solution is expressed in terms of elementary functions and is much simpler than even the starting expression given by Eq.~(\ref{standard_spectrum}). 

The Gaussian distribution of a random field is a well-known situation in physics (see, e.g., \cite{RueHeld}). Such a distribution for local magnetic field strengths is called upon in the literature to explain, for example, properties of synchrotron radiation from  supernova remnants \citep{GaussianSNR} and pulsar wind nebulae \citep{GaussianPWN}. Recently, in the context of synchrotron emission of supernova remnants, this issue has been discussed also by \cite{Pohl2015}. Note that 
in this paper the authors assumed Gaussian distribution for the total magnetic field strength, unlike the usual approach where all three magnetic field components are Gaussian-distributed. 

The Gaussian distribution naturally results from summation of the magnetic field from many independent or nearly independent modes, for example in frequently occurring case of quasi-linear turbulence. Isotropic or nearly isotropic distribution of electrons over pitch angles is expected in the case where the cooling length exceeds the electrons' mean free path, that is also a typical situation.

Since our results are derived from Eq.~(\ref{standard_spectrum}), they inherit all its limitations, which are to be applied to the typical magnetic field strength. In addition we require that the radiating particles effectively isotropize over a distance smaller than their cooling length.

The paper is organized as follows. In Section~\ref{sec:fieldDistribution} we discuss how distribution of electrons over pitch angles can be convoluted with the distribution of local magnetic field strength to obtain the effective magnetic field distribution and derive this distribution for the case of Gaussian turbulence. In Section~\ref{sec:individualSySpectrum} we derive the synchrotron spectrum of an individual electron, averaged over pitch angles and over the magnetic field strength distribution. The steps required to evaluate the integral, which expresses this spectrum, are outlined in Section~\ref{sec:appendix1}. We move on obtaining in Section~\ref{sec:PL_spectrum} expression for the spectrum of synchrotron radiation produced in turbulent magnetic field by electrons, whose distribution function is a power-law with cut-off. We also derive, in Section~\ref{sec:uncut_PL}, the spectrum from simple (uncut) power-law distribution and elaborate on its relation to the spectrum derived in monochromatic approximation.
In Section~\ref{sec:comparison} we compare our expressions for synchrotron spectrum in turbulent magnetic fields to those for the uniform magnetic field.

\section{Effective distribution of magnetic field strength for Gaussian turbulence} 
\label{sec:fieldDistribution}

We assume that the local strength of magnetic field in the emitting region results from summation of many independent modes, that means independent Gaussian distribution with zero mean value for all three Cartesian components of the field. Then the probability density for the magnetic field strength is 
\begin{equation} \label{field_distribution}
\begin{array}{l}
    \displaystyle
    P_B = \left(\frac{6}{\pi}\right)^{1/2} \frac{3 B^2}{B_0^3} \exp{\left(-\frac{3 B^2}{2 B_0^2}\right)}\, ;\\[3ex]
    \displaystyle
    \int_0^\infty P_B(B)\, {\rm d}B = 1\, , \qquad \langle B^2 \rangle  = \int_0^\infty B^2 P_B(B)\, {\rm d}B = B_0^2 \, .
\end{array}
\end{equation}
This assumption is natural in the case, where turbulence is sustained in quasi-linear regime, and we consider it a conservative (i.e., underestimating the volume occupied by stronger than average magnetic field) assumption in the case of strongly non-linear turbulence.

A particle moving along helical trajectory with pitch angle $\theta \gg 1/\gamma$ behaves (in terms of radiated power and spectrum) as if it moves perpendicular to field lines of the magnetic field with effective strength $B^\mathrm{eff}(\theta)= B \sin \theta$. In case where particle distribution over pitch angles does not vary from point to point, there are two ways to calculate the average synchrotron spectrum. One may either integrate locally over pitch angles and then average over the field strength distribution or -- equivalently -- calculate effective field strength distribution for all particles with the same pitch angle and after that integrate over pitch angles. We follow the second route.

Effective field strength distribution is formally the same as Eq.~\ref{field_distribution}, where $B$ is replaced by $B^\mathrm{eff}(\theta)= B \sin \theta$ and $B_0$ is replaced by $B_0^{\rm eff}(\theta) = B_0 \sin\theta$.
Isotropic distribution over pitch angles is equivalent to the following distribution over $B_0^{\rm eff}$,
\begin{equation}
    P_{\theta} = \left\lvert \frac{\partial \cos\theta}{\partial B_0^{\rm eff}} \right\rvert
    = \left\lvert \frac{\partial}{\partial B_0^{\rm eff}} \sqrt{1-\left(\frac{B_0^{\rm eff}}{B_0}\right)^2} \right\rvert
    = \frac{B_0^{\rm eff}}{B_0 \sqrt{B_0^2-(B_0^{\rm eff})^2}}
    \, ,\\
\end{equation}
resulting in 
\begin{equation} \label{effective_P_B}
\begin{array}{l}
    \displaystyle
    P_B^{\rm eff} = \int_0^{B_0} P_B^\mathrm{eff}\! (\theta) P_{\theta}\, {{\rm d} B_0^{\rm eff}} 
    = \frac{3 B^\mathrm{eff}}{B_0^2} \exp{\left(-\frac{3 (B^\mathrm{eff})^2}{2 B_0^2}\right)}\, ;\\[3ex]
    \displaystyle
    \int_0^{\infty} P_B^{\rm eff}\, {{\rm d} B^{\rm eff}} = 1\, \qquad 
    \langle (B^{\rm eff})^2 \rangle = \frac{2}{3} B_0^2   \, .
\end{array}
\end{equation}
We perform this integration by making substitution $\displaystyle u= \frac{3}{2} (B^\mathrm{eff})^2 \left( \frac{1}{(B_0^{\rm eff})^2} - \frac{1}{B_0^2} \right)$. After that, the integration with respect to $u$ results in gamma function, $\Gamma(1/2) = \sqrt{\pi}$.

\section{Average synchrotron spectrum for isotropic particles in turbulent magnetic field} 
\label{sec:individualSySpectrum}

For our purposes it is more convenient to express the synchrotron spectrum (Eq.~\ref{standard_spectrum}) in terms of the number of synchrotron photons emitted per unit frequency, which is
\begin{equation} \label{Nsy1}
    \left( \frac{{\rm d} \dot{N}}{{\rm d} \omega} \right)_u = \frac{1}{\hbar \omega}\, \frac{{\rm d} L}{{\rm d} \omega}
    = \frac{1}{\gamma^2}\,  \frac{\alpha}{3}\,  N\!\left( \frac{\omega}{\omega_c} \right)\, .
\end{equation}
Here $\alpha$ is the fine-structure constant and
\begin{equation}
    N(x) = \frac{\sqrt{3}}{\pi}\,  \frac{1}{x}\, F(x) = \frac{\sqrt{3}}{\pi}  \int_x^\infty K_{5/3}(\xi) \, {\rm d} \xi \, .
\end{equation}
Numerical factor in definition of $N(x)$ is chosen to simplify further notation.

Using effective distribution of the magnetic field (Eq.~\ref{effective_P_B}) instead of actual one, 
we treat all particles as if they were moving perpendicular to the field lines, so that the distribution of synchrotron photons over frequency, averaged over space and pitch angles, is
\begin{equation} \label{Eq}
    \left( \frac{{\rm d} \dot{N}}{{\rm d} \omega} \right)_t 
    = \int_0^\infty P_B^{\rm eff}(B) \left[
 \frac{1}{\gamma^2} \frac{\alpha}{3}  N\!\left( \frac{2}{3}\, \frac{m_e c\, \omega}{\gamma^2 e B} \right)    
 \right] \, {\rm d} B  \, ,\\
\end{equation}
where subscript $t$ denotes turbulent magnetic field.

Noting that $P_B^{\rm eff}(B) {\rm d} B$ is the exact differential, and   integrating Eq.~(\ref{Eq}) by parts using the  
substitution 
\begin{equation*}
\xi = \frac{8}{9}\, \frac{\omega B_0}{\omega_0 B}\, , \qquad  
\omega_0 = \frac{4}{3} \gamma^2 \frac{eB_0}{m_e c} \equiv  \frac{4}{3} \gamma^2 \omega_\mathrm{_{B,0}}\, ,
\end{equation*}
(here and below we use the equivalent of $\omega_\mathrm{_B}$ for the turbulent magnetic field, $\omega_\mathrm{_{B,0}} = \frac{eB_0}{m_e c}$.)
we arrive  at 
\begin{multline} \label{turbylentSy}
    \left( \frac{{\rm d} \dot{N}}{{\rm d} \omega} \right)_t
    =  \frac{1}{\gamma^2} \frac{\alpha}{3}  
    \int_0^\infty 
    N\!\left( \frac{8}{9}\, \frac{\omega}{\omega_0}\, \frac{B_0}{B} \right) 
    \left[ \frac{3 B}{B_0^2} \exp{\left(-\frac{3 B^2}{2 B_0^2}\right)} \right]
    \, {\rm d} B  \\
    =  \frac{1}{\gamma^2} \frac{\alpha}{\pi \sqrt{3}}\,   
    \frac{8\omega}{9\omega_0}\int_0^\infty 
    \frac{B_0}{B^2}\, \exp{\left(-\frac{3 B^2}{2 B_0^2}\right)}\,
    K_{5/3}\! \left( \frac{8}{9}\, \frac{\omega}{\omega_0}\, \frac{B_0}{B} \right)
    \, {\rm d} B  \\
    =  \frac{1}{\gamma^2} \frac{\alpha}{\pi \sqrt{3}}\,   
    \int_0^\infty 
    \exp{\left(-\frac{32}{27}\, \frac{(\omega/\omega_0)^2}{\xi^2}\right)}\,
    K_{5/3}\! \left( \xi \right)
    \, {\rm d} \xi  \\
    =  \frac{1}{\gamma^2} \frac{\alpha}{3}\,  Q\! \left(\frac{\omega}{\omega_0}\right)  \, ,  
\end{multline}
where 
\begin{equation} \label{Q_definition}
    Q(x) = \frac{\sqrt{3}}{\pi} \int_0^\infty \exp{\left(-\frac{32 x^2}{27 \xi^2}\right)}\, K_{5/3}(\xi) \, {\rm d} \xi 
    = \left( 1  + \frac{1}{x^{2/3}} \right)  \exp \left( - 2 x^{2/3} \right)  
\end{equation}
is evaluated in Sect.~\ref{sec:appendix1}.
Note that the numerical factor (4/3) in definition of $\omega_0$ differs from the numerical factor (3/2) in definition of $\omega_c$. This seemingly cumbersome choice is made to simplify notation in the final expression.

\section{Synchrotron spectrum from power-law distribution with cut-off}
\label{sec:PL_spectrum}

A rather general approximation for distribution function of synchrotron-radiating particles is power-law with cut-off,
\begin{equation} \label{PLdistribution}
    f_e \left( \gamma \right)
    = \frac{n_e}{\gamma_0} \left( \frac{\gamma}{\gamma_0} \right)^{-p} \, \exp{\left( - \left[ \frac{\gamma}{\gamma_0} \right]^\beta \right)}\, ,
\end{equation}
where $\gamma_0 \gg 1$ and $\beta>0$. We are only interested in the part of this distribution, where $\gamma \gg 1$, so that we can formally assume that the distribution (\ref{PLdistribution}) extends to $\gamma=0$; this simplifies notation. Not that we do not require the integral $\int_0^\infty f_e \, {\rm d} \gamma$ to converge.

Calculation of spectral distribution of synchrotron photons for power-law distribution of radiating particles yields
\begin{multline} \label{PL_spectrum}
    \frac{{\rm d} \dot{N}_{\rm PL}}{{\rm d} \omega}(\omega, B_0) 
    = \int_{0}^\infty \frac{{\rm d} \dot{N}}{{\rm d} \omega} \left( \omega, \gamma, B_0 \right) \, f_e \, {\rm d} \gamma \\
    = \frac{\alpha}{3}\, \frac{n_e}{\gamma_0^2}  \int_{0}^\infty Q\! \left(\frac{\omega}{\omega_\mathrm{cut}} \frac{\gamma_0^2}{\gamma^2}\right) 
    \left( \frac{\gamma}{\gamma_0} \right)^{-2-p} \, \exp{\left( - \left[ \frac{\gamma}{\gamma_0} \right]^\beta \right)} 
    \, \frac{{\rm d} \gamma}{\gamma_0} \\
    = \frac{\alpha}{3}\, \frac{n_e}{\gamma_0^2}  \left( \frac{\omega}{\omega_\mathrm{cut}} \right)^{-\frac{(1+p)}{2}} 
    \int_{0}^\infty Q\! \left(\frac{1}{\xi^2}\right) 
    \xi^{-2-p} \, \exp{\left( - \left[ \frac{\omega}{\omega_\mathrm{cut}} \right]^{\beta/2}  \xi^\beta \right)}   \, {\rm d} \xi \\
    = \frac{\alpha}{3}\, \frac{n_e}{\gamma_0^2}\,  Q_\mathrm{PL}\! \left(\frac{\omega}{\omega_\mathrm{cut}}\right) \, .
\end{multline}
Here we changed integration variable to $\displaystyle \xi = \sqrt{\frac{\omega_\mathrm{cut}}{\omega}} \frac{\gamma}{\gamma_0}$, and introduced
\begin{equation}
    \omega_\mathrm{cut} = \frac{4}{3} \gamma_0^2 \omega_\mathrm{_{B,0}} \, .
\end{equation}
Substituting $Q$ from Eq.~(\ref{Q_definition}), we write the function $Q_\mathrm{PL}(x)$ explicitly:
\begin{equation} \label{Q_PL}
    Q_\mathrm{PL}(x)
    = x^{-\frac{(1+p)}{2}}  \int_{0}^\infty 
    \left( \xi^{-p-2}  + \xi^{-p-2/3} \right)  
    \, \exp{\left(  - 2 \xi^{-4/3} - x^{\beta/2}  \xi^\beta \right)}   \, {\rm d} \xi \, ,
\end{equation}
where
\begin{equation}
    x = \frac{\omega}{\omega_\mathrm{cut}} 
    = \frac{3}{4} \frac{\omega}{\gamma_0^2 \omega_\mathrm{_{B,0}}}   \, .
\end{equation}

For practical purposes it is useful to derive asymptotic forms of the function $Q_\mathrm{PL}(x)$ as well as its approximation in terms of elementary functions.
Evaluating the asymptotic form in the limit $x \rightarrow \infty$ we note, that the main contribution to the integral in Eq.~(\ref{Q_PL}) comes from $\xi \rightarrow 0$, keep only the smallest power $\xi$, and then use Laplace's method (see Sect.~\ref{sec:appendix2}).
Here $g(\xi)= \xi^{-p-2}$ 
and $f(\xi)=   - 2 \xi^{-4/3} - x^{\beta/2}  \xi^\beta$, so that 
\begin{equation}
\begin{array}{rcl}
    \xi_0  & = & 
    \displaystyle
    \left( \frac{8}{3 \beta  x^{\beta/2}} \right)^{\frac{3}{3\beta+4}}\\[3ex]
    \lvert f^{\prime \prime}(\xi_0) \rvert & = & 
    \displaystyle
    \frac{8}{3} \left( \frac{3 \beta  x^{\beta/2}}{8} \right)^{\frac{10}{3\beta+4}} \left(\beta + \frac{4}{3}  \right)\\[3ex]
    f(\xi_0) & = &     
    \displaystyle
    - \left( \frac{3 \beta  x^{\beta/2}}{8} \right)^{\frac{4}{3\beta+4}} \left(2 + \frac{8}{3 \beta} \right)     \\
\end{array}
\end{equation}
and the asymptotic form at large arguments is
\begin{equation}
    Q_\mathrm{PL}(x) 
    = \frac{3 \sqrt{\pi}}{2 \left(3\beta + 4 \right)^{1/2}} \,
    \left( \frac{3 \beta}{8} \right)^{\frac{3p+1}{3\beta+4}} 
    x^{- \frac{ \beta + 2  + 2p}{3\beta+4}}  
    \exp{\left[ - \left(2 + \frac{8}{3 \beta} \right) \left( \frac{3 \beta}{8} \right)^{\frac{4}{3\beta+4}} x^{\frac{2\beta}{3\beta+4}} \right]}   \, .
\end{equation}

For the asymptotic form of the function $Q_\mathrm{PL}(x)$ in the limit $x \rightarrow 0$ there are three cases depending on the value of $p$:
\begin{equation}
    Q_\mathrm{PL}(x) 
    = x^{-\frac{(1+p)}{2}} \times  \left\{
    \begin{array}{ll}
        I_1\! \left( -p-2 \right) + I_1\! \left( -p-2/3 \right) \, , & \qquad p > 1/3  \\[1ex]
        I_2 \, , & \qquad p = 1/3  \\[1ex]
        I_3\! \left( -p-2/3 \right) \, , & \qquad p < 1/3   \, ,
    \end{array}
    \right.
\end{equation}
where 
\begin{equation} \label{case1}
    I_1\! \left( q \right) 
    =  \int_{0}^\infty \xi^{q} \, \exp{\left(  - 2 \xi^{-4/3} \right)}   \, {\rm d} \xi
    = 3 \times 2^{\frac{3 q -5 }{4}}\, \Gamma\! \left( -\frac{3}{4} (1 + q) \right) \, ,
\end{equation}
\begin{multline} \label{case2}
    I_2 
    =  \int_{0}^\infty \xi^{-1} \, \exp{\left(  - 2 \xi^{-4/3} - x^{\beta/2}  \xi^\beta \right)}   \, {\rm d} \xi \\
    =  \int_{0}^{A} \xi^{-1} \, \exp{\left(  - 2 \xi^{-4/3} \right)}   \, {\rm d} \xi 
    +  \int_{A}^\infty \xi^{-1} \, \exp{\left(  - x^{\beta/2}  \xi^\beta \right)}   \, {\rm d} \xi \\
    = -\frac{3}{4}\, \mathrm{Ei}\! \left( -\frac{2}{A^{4/3}} \right)
    -  \frac{1}{\beta} \, \mathrm{Ei}\! \left( -x^{\beta/2}  A^\beta \right) 
    \simeq  -  \frac{1}{4} \ln{\left( 8 x^2 \right) }     -   \left( \frac{3}{4} + \frac{1}{\beta} \right) \gamma_{_\mathrm{Eu}}
\end{multline}
and
\begin{equation} \label{case3}
    I_3\! \left( q \right) 
    = \int_{0}^\infty \xi^{q} \, \exp{\left(  - x^{\beta/2}  \xi^\beta \right)}   \, {\rm d} \xi 
    = \frac{1}{\beta}\, x^{-(q+1)/2}\, \Gamma\! \left( \frac{q+1}{\beta}  \right)     \, .
\end{equation}
The integrals in Eqs.~(\ref{case1}) and (\ref{case3}) are reduced to gamma function by substitutions $\zeta = 2 \xi^{-4/3}$ and $\zeta = x^{\beta/2}  \xi^\beta$ (i.e., $\xi  = \zeta^{1/\beta} x^{-1/2}$), respectively. To evaluate the integral in Eq.~(\ref{case2}), we first split it into two parts, integrating from 0 to $A$ and from $A$ to infinity; $1 \ll A \ll x^{-1/2}$. In the first part we keep only the cutoff at $\xi < 1$ and in the second part we keep only the cutoff at $\xi \rightarrow \infty$. With 
substitutions of integration variable ($\zeta = 2 \xi^{-4/3}$ in the first part and $\zeta = x^{\beta/2}  \xi^\beta$ in the second part), both parts are reduced to exponential integrals. Then we use the asymptotic form at $x \rightarrow 0$,  $\mathrm{Ei}\! \left( x \right) = \ln{(x)} + \gamma_{_\mathrm{Eu}}$ (the first two terms from Puiseux series), where $\gamma_{_\mathrm{Eu}} \simeq 0.5772156649$  is the Euler-Mascheroni constant, to obtain the final result.

The asymptotic forms of the function $Q_\mathrm{PL}(x)$ can be summarized as follows. For small frequencies (in the limit $x \rightarrow 0$)
\begin{equation} \label{asymp_at_small_x}
    Q_\mathrm{PL}(x) 
    =   \left\{
    \begin{array}{ll}
        C_1 x^{-\frac{(1+p)}{2}}  \, , & \qquad p > 1/3  \\[1ex]
        \left[ -  \frac{1}{4} \ln{\left( 8 x^2 \right) }     -   C_2  \right] x^{-2/3}
        \, , & \qquad p = 1/3  \\[1ex]
        C_3 x^{-2/3}  \, , & \qquad p < 1/3   \, ,
    \end{array}
    \right.
\end{equation}
where 
\begin{equation} 
    C_1 
    = 2^{-\frac{3p+7}{4}}\, \frac{3 \left( 3 p + 7 \right)}{8} \, \Gamma\! \left( \frac{3 p - 1}{4}  \right)     \, ,
\end{equation}
\begin{equation} 
    C_2     = \left( \frac{3}{4} + \frac{1}{\beta} \right) \gamma_{_\mathrm{Eu}}    \, ,
\end{equation}
\begin{equation} 
    C_3 = \frac{1}{\beta}  \Gamma\! \left( \frac{1 - 3 p}{3\beta}  \right)    \, .
\end{equation}
 For large frequencies (in the limit $x \rightarrow \infty$)
\begin{equation} \label{asymp_at_large_x}
    Q_\mathrm{PL}(x)     = C_4\,   x^{- \frac{ \beta + 2  + 2p}{3\beta+4}}  
    \exp{\left( - C_5\, x^{\frac{2\beta}{3\beta+4}} \right)}  \, ,
\end{equation}
where
\begin{equation}
    C_4     = \frac{3 \sqrt{\pi}}{2 \left(3\beta + 4 \right)^{1/2}} \,
    \left( \frac{3 \beta}{8} \right)^{\frac{3p+1}{3\beta+4}}    \, ,
\end{equation}
\begin{equation}
    C_5 =   \left(2 + \frac{8}{3 \beta} \right) \left( \frac{3 \beta}{8} \right)^{\frac{4}{3\beta+4}}      \, .
\end{equation}

\begin{figure}
    \centering
    \includegraphics[width=1.0\columnwidth]{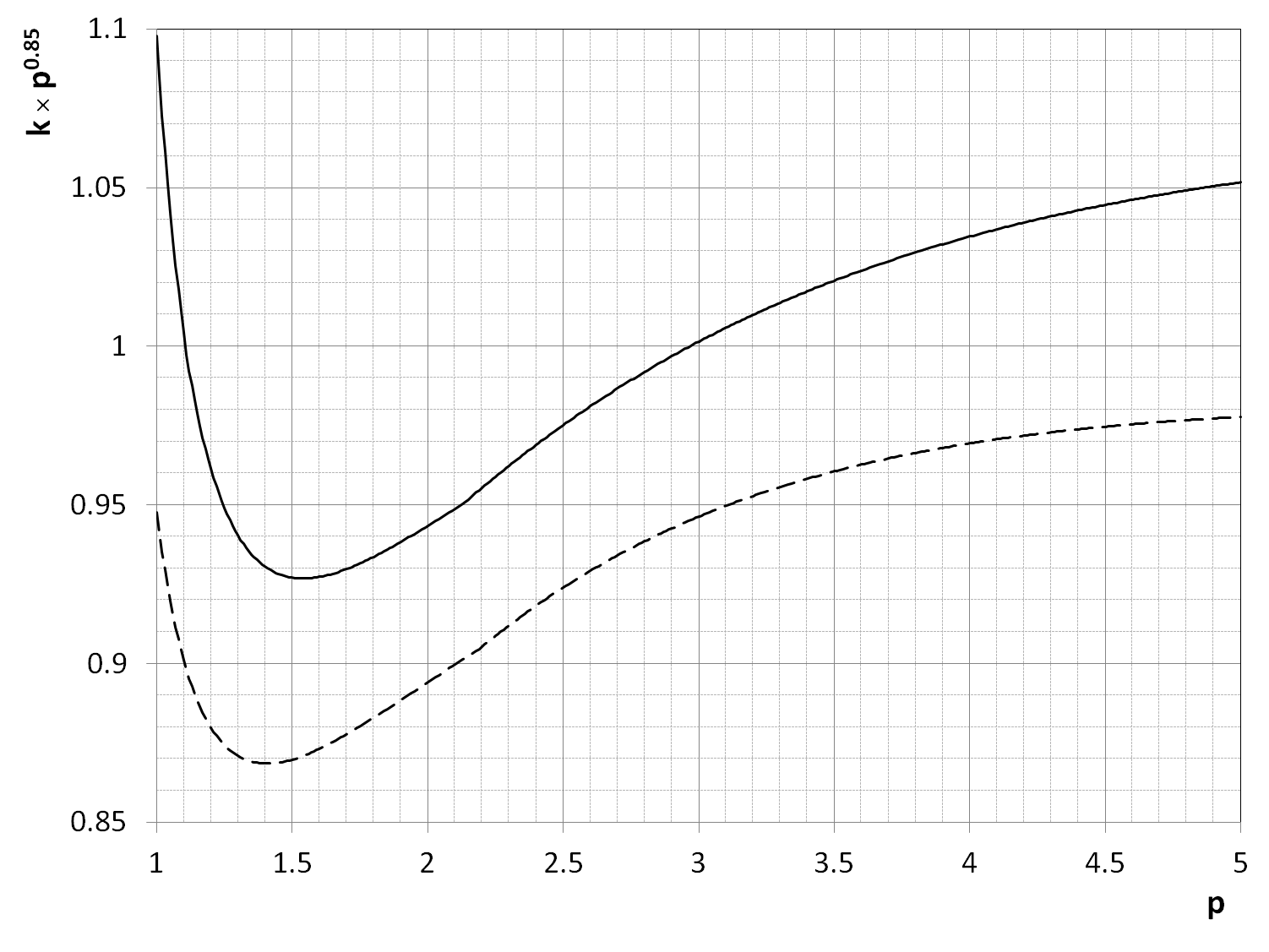}
    \caption{The best-fit value of $k$-parameter in approximation given by Eq.(\ref{two_sided_approximation}) for $\beta=2$ (solid line) and $\beta=1$ (dashed line). Vertical axis -- $k p^{0.85}$. Horizontal axis -- power-law index $p$.}
    \label{k_parameter}
\end{figure}

\begin{figure}
    \centering
    \includegraphics[width=1.0\columnwidth]{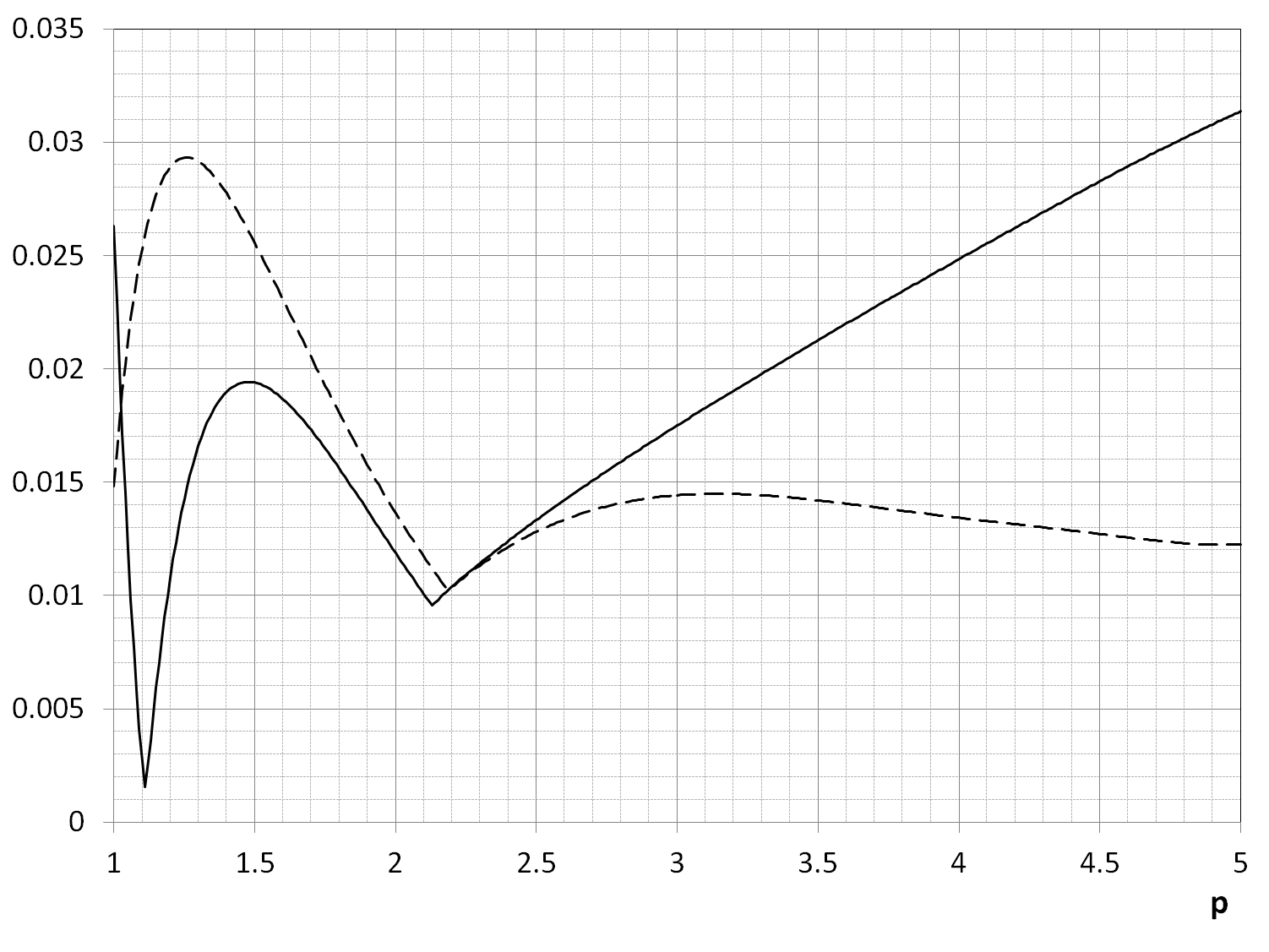}
    \caption{The largest relative error for the synchrotron spectrum fit given by Eq.~(\ref{two_sided_approximation}) with $k$-parameter chosen according to Fig.~\ref{k_parameter}. Solid line -- $\beta=2$, dashed line -- $\beta=1$.  Vertical axis -- the error, horizontal axis -- power-law index $p$.}
    \label{fit_errors}
\end{figure}

Combining the two asymptotic forms given by Eqs.(\ref{asymp_at_small_x}) and (\ref{asymp_at_large_x}) one obtains an approximation, that is valid for any $x$. For example,  in the case $p>1/3$, which covers the vast majority of situations relevant to astrophysics, we arrive at the following approximate and asymptotically exact expression:
\begin{equation} \label{two_sided_approximation}
    Q_\mathrm{PL}(x) \simeq \left( t_1^k  + t_2^k  \right)^{1/k}
    \exp{\left( - C_5\, x^{\frac{2\beta}{3\beta+4}} \right)}  \, .
\end{equation}
where 
\begin{equation} 
    t_1 = C_1\, x^{-\frac{(1+p)}{2}}  \qquad \mbox{and} \, \qquad
    t_2 = C_4\, x^{- \frac{ \beta + 2  + 2p}{3\beta+4}}  
\end{equation}
and $k$ is the parameter whose value is chosen to minimize error for each ($p$,$\beta$) pair. 
For $\beta=1,2$ and arbitrary power-law index $p$ the value of $k$-parameter can be taken from Fig.~(\ref{k_parameter}) and the largest relative error is plotted in Fig.~\ref{fit_errors}.

%--------------------Fig1

%--------------------------Fig.2

\section{Synchrotron spectrum for simple power-law distribution and connection to monochromatic approximation}
\label{sec:uncut_PL}

At times one is interested only in the low-energy part of distribution (\ref{PLdistribution}), that can be approximated by a simple power-law,
\begin{equation} \label{uncutPL}
    f_e \left( \gamma \right)
    = \frac{n_e}{\gamma_0} \left( \frac{\gamma}{\gamma_0} \right)^{-p} \, .
\end{equation}
The synchrotron spectrum for this distribution is given by the expression similar to Eq.~(\ref{Q_PL}), but without second term in the exponent,
so that Eq.~(\ref{asymp_at_small_x}) is exact expression for $Q_\mathrm{PL}(x)$ rather than its asymptotic form in the limit $x \rightarrow 0$.
Considering once again typical in astrophysics case $p>1/3$, we obtain
\begin{equation}  \label{uncutPL_exact}
    \frac{{\rm d} \dot{N}_{\rm PL}}{{\rm d} \omega}(\omega, B_0) 
    =   \frac{\alpha}{8}\, \frac{n_e}{\gamma_0^2}\,  2^{-\frac{3p+7}{4}}\, \left( 3 p + 7 \right) \, \Gamma\! \left( \frac{3 p - 1}{4}  \right) 
    \left(\frac{\omega}{\omega_\mathrm{cut}}\right)^{-\frac{p+1}{2}}     \, .
\end{equation}
The synchrotron spectrum obtained for a power-law distribution in monochromatic approximation (i.e., assuming that each electron radiates at a single frequency $\omega_m$, proportional to the square of its Lorentz factor) has the same frequency dependence. To ensure that the numerical factor is also the same it is necessary to choose the frequency $\omega_m$ in appropriate way:
\begin{equation} \label{omega_m}
    \omega_m = a_m \gamma^2 \, \omega_\mathrm{_{B,0}} 
\end{equation}
and
\begin{equation} \label{a_m}
     a_m    =  \frac{\sqrt{2}}{3} \left[  \frac{3 p + 7}{16} \, \Gamma\! \left( \frac{3 p - 1}{4}  \right)  \right]^{\frac{2}{p-3}}  \, .
\end{equation}
Note that $a_m(p)$ is monotonously rising function of $p$; it equals zero at $p=1/3$ and is continuous at $p=3$.

\section{Discussion} 
\label{sec:comparison}

It is instructive to compare spectra of synchrotron radiation, the total emitted power and the average energy of synchrotron photons in two cases: for an electron moving perpendicular to the field lines of uniform magnetic field and for an electron in the turbulent magnetic field, which has the same average energy density.

For an electron moving perpendicular to the field lines of uniform magnetic field, the total emitted power is
\begin{multline} \label{SyLuminocity_uniformB}
   L_u = \int_0^\infty \hbar \omega \left( \frac{{\rm d} \dot{N}}{{\rm d} \omega} \right)_u  {\rm d} \omega 
   =  \hbar \omega_c^2 \, \frac{1}{\gamma^2} \frac{\alpha}{3} \int_0^\infty  x\, N(x)\, {\rm d} x    
   =  \frac{8}{27} \,  \alpha \, \frac{\hbar \omega_c^2}{\gamma^2}   \\
   = 2 \gamma^2 \sigma_T c \frac{B_0^2}{8\pi}
\end{multline}
and the total photon emission rate is
\begin{multline} \label{SyPhotonRate_uniformB}
   \dot{N}_u = \int_0^\infty \left( \frac{{\rm d} \dot{N}}{{\rm d} \omega} \right)_u  {\rm d} \omega 
   = \omega_c \, \frac{1}{\gamma^2} \,  \frac{\alpha}{3} \int_0^\infty  N\! \left(x\right) \, {\rm d} x
   = \frac{5}{3\sqrt{3}} \, \frac{\alpha \omega_c}{\gamma^2}  \\
   = \frac{5}{2\sqrt{3}} \, \alpha \omega_B\, . 
\end{multline}
To find the integrals
\begin{equation*}
     \int_0^\infty  N(x)\, {\rm d} x 
     = \frac{\sqrt{3}}{\pi}  \int_0^\infty  x K_{5/3}(x) \, {\rm d} x
     = \frac{\sqrt{3}}{\pi} \, \Gamma\!\left(\frac{1}{6}\right) \Gamma\!\left(\frac{11}{6}\right)
     = \frac{5}{\sqrt{3}}
\end{equation*}
and 
\begin{equation*}
     \int_0^\infty  x\, N(x)\, {\rm d} x 
     = \frac{\sqrt{3}}{2 \pi}  \int_0^\infty  x^2 K_{5/3}(x) \, {\rm d} x
     = \frac{\sqrt{3}}{\pi} \, \Gamma\!\left(\frac{2}{3}\right) \Gamma\!\left(\frac{7}{3}\right)
      = \frac{8}{9}
\end{equation*}
we first integrate by parts, then use general expression (see, e.g., Eq.~6.561.16 in \cite{GradshteynRyzhik})
\begin{equation*}
    \int_{0}^{\infty}x^{\mu}K_{\nu}\left(ax\right) \, \mathrm{d}x 
    = 2^{\mu-1} \, a^{-\mu-1} \, \Gamma\!\left(\frac{1+\mu-\nu}{2}\right) \Gamma\!\left(\frac{1+\mu+\nu}{2}\right)
\end{equation*}
and Euler's reflection formula
\begin{equation*}
    \Gamma\left(z\right)\Gamma\left(1-z\right)=\frac{\pi}{\sin\left(\pi z\right)} \, .
\end{equation*}

From Eqs.(~\ref{SyLuminocity_uniformB}) and (\ref{SyPhotonRate_uniformB}) we find the average energy of synchrotron photons emitted by an electron moving perpendicular to the field lines of uniform magnetic field:
\begin{equation} \label{AvgSyEnergy_uniformB}
   \langle \epsilon_{\rm sy} \rangle_u = L_u/\dot{N}_u 
   = \frac{8}{15\sqrt{3}} \, \hbar \omega_c 
   = \frac{4}{5\sqrt{3}} \, \gamma^2 \hbar \omega_B \, .
\end{equation}

Similarly, for an electron in the turbulent magnetic field, the total emitted power is
\begin{multline} \label{SyLuminocity_turbulentB}
   L_t = \int_0^\infty \hbar \omega \left( \frac{{\rm d} \dot{N}}{{\rm d} \omega} \right)_t {\rm d} \omega 
   =  \hbar \omega_0^2 \, \frac{1}{\gamma^2} \frac{\alpha}{3} \int_0^\infty  x\, Q(x)\, {\rm d} x    
   =  \frac{\alpha}{4} \, \frac{\hbar \omega_0^2}{\gamma^2}   \\
   = \frac{4}{3} \gamma^2 \sigma_T c \frac{B_0^2}{8\pi} 
\end{multline}
and the total photon emission rate is
\begin{multline} \label{SyPhotonRate_turbulentB}
   \dot{N}_t = \int_0^\infty \left( \frac{{\rm d} \dot{N}}{{\rm d} \omega} \right)_t {\rm d} \omega 
   = \omega_0 \, \frac{1}{\gamma^2} \,  \frac{\alpha}{3} \int_0^\infty  Q\! \left(x\right) \, {\rm d} x
   = \frac{5}{8} \sqrt{\frac{\pi}{2}} \, \frac{\alpha \omega_0}{\gamma^2} \\
   = \frac{5}{6} \sqrt{\frac{\pi}{2}} \, \alpha \omega_\mathrm{_{B,0}}\, . 
\end{multline}
The integrals
\begin{equation*} 
    \int_0^{\infty} Q(x) \, {\rm d} x   = \frac{15}{16} \sqrt{2 \pi} 
    \qquad {\rm and} \qquad
    \int_0^{\infty} x\, Q(x) \, {\rm d} x   = \frac{3}{4} 
\end{equation*}
are calculated in a straightforward way (reduced to the gamma function).

The average energy per synchrotron photon,
\begin{equation}  \label{AvgSyEnergy_turbulentB}
   \langle \epsilon_{\rm sy} \rangle_t = L_t/\dot{N}_t 
   = \frac{2}{5} \, \sqrt{\frac{2}{\pi}} \; \hbar \omega_0  
   =  \frac{8}{15} \sqrt{\frac{2}{\pi}} \gamma^2 \hbar \omega_\mathrm{_{B,0}} \, ,
\end{equation}
is approximately equal ($\simeq 1.0854$ times smaller) to the value in Eq.~(\ref{AvgSyEnergy_uniformB}). 

Here we may note that the net effect of turbulent magnetic field is to increase the average energy of synchrotron photons by a factor, which approximately compensates decrease of this energy due to averaging over isotropic pitch angle distribution. The synchrotron luminosity for an electron in the turbulent is 2/3 of the value given by Eq.~(\ref{SyLuminocity_uniformB}). This difference is due to the fact that one of three components of the turbulent magnetic field (the one parallel to electron's momentum) does not contribute to synchrotron radiation. 
The same factor appears when Eq.~(\ref{SyLuminocity_uniformB}) is averaged over isotropic pitch angle distribution.

%---------------------Fig.3
\begin{figure}
    \centering
    \includegraphics[width=1.0\columnwidth]{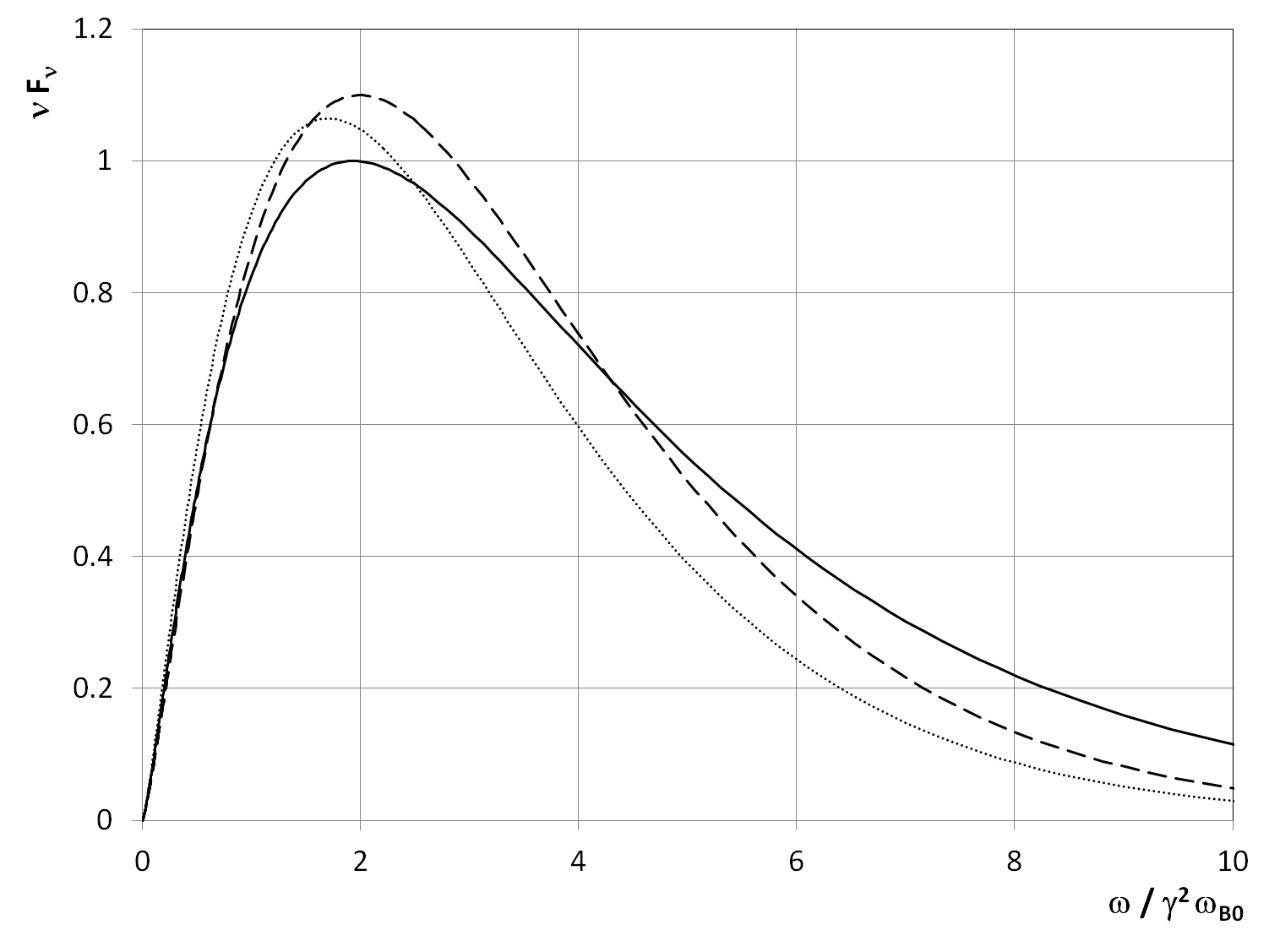}
    \caption{Spectral energy distributions of synchrotron radiation for an electron in the turbulent magnetic field (solid line) and in the uniform magnetic field, averaged over isotropic pitch angle distribution (dotted line) and for $\pi/2$ pitch angle (dashed line). The latter SED is multiplied by 2/3 to have the same total radiated power. Vertical axis -- $\nu F_\nu$ in arbitrary units. Horizontal axis -- frequency.}
    \label{SySED_LinearScale}
\end{figure}

The synchrotron spectral energy distributions (SEDs) for mono-energetic electrons in turbulent and uniform magnetic fields are compared in Fig.~\ref{SySED_LinearScale}. 
Note that for the uniform magnetic field the SED of electrons moving perpendicular to the field lines peaks at $\omega \simeq 1.99\, \gamma^2 \omega_\mathrm{_{B,0}}$, while isotropic population of electrons in the same field produces SED which peaks at  $\omega \simeq 1.70\, \gamma^2 \omega_\mathrm{_{B,0}}$. 
Presence of regions with a stronger field in the case of Gaussian turbulent magnetic field almost exactly compensates the decrease of the SED peak frequency due to averaging over isotropic pitch angle distribution, so that the turbulent-field SED peaks at $\omega \simeq 1.93\, \gamma^2 \omega_\mathrm{_{B,0}}$. 

The spectra 
of synchrotron radiation in the cases of turbulent and uniform magnetic fields 
are rather similar at low frequencies, below and around the peak, but the difference between them becomes progressively larger at high frequencies. Although the difference exceeds factor of 2 only at the highest frequencies, where $\simeq 1.5$ per cent of emitted power is concentrated, it shows up in electron distributions with sharp high-energy cut-off.

\begin{figure}
    \centering
    \includegraphics[width=0.67\columnwidth]{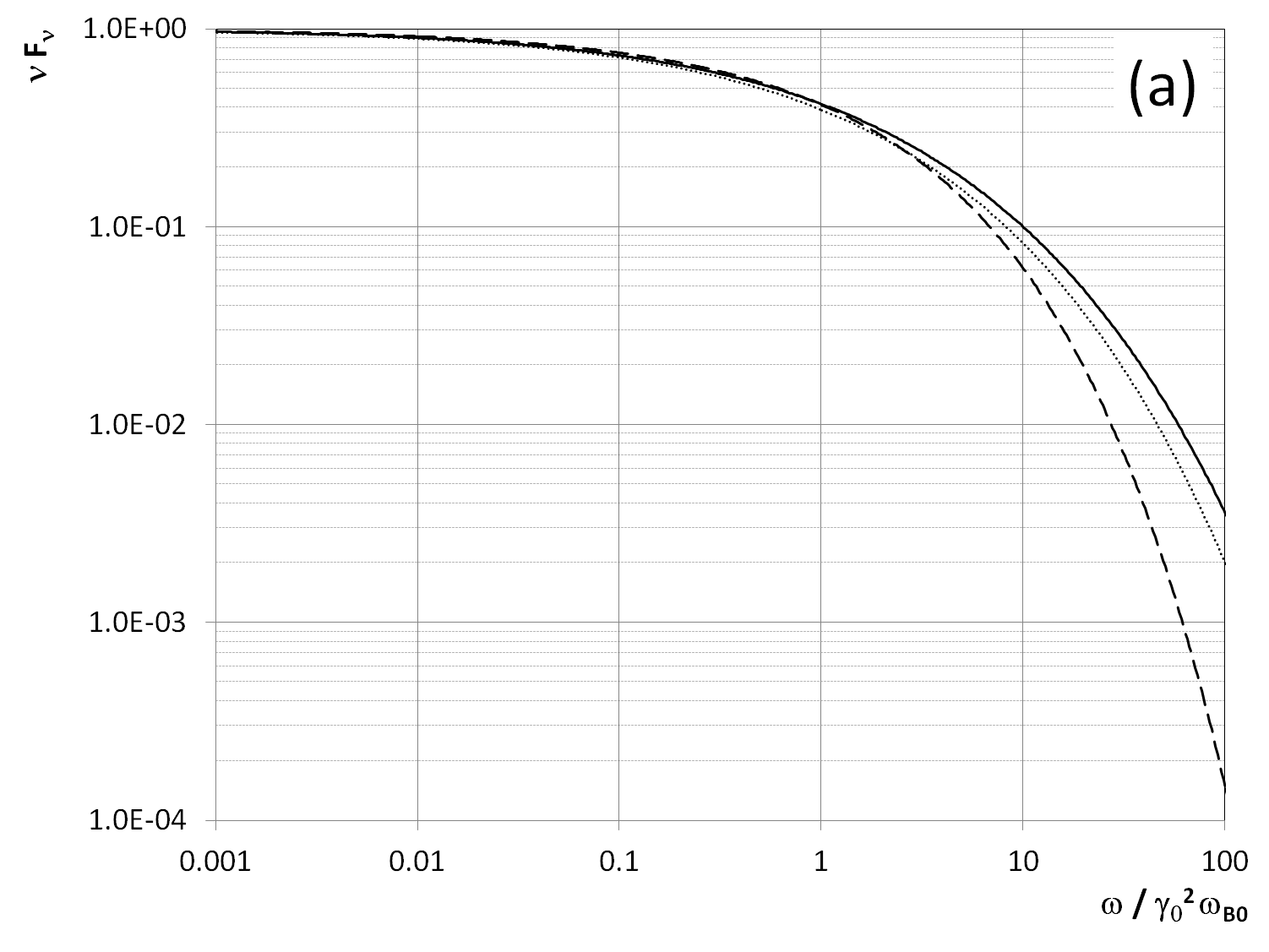}
    \includegraphics[width=0.67\columnwidth]{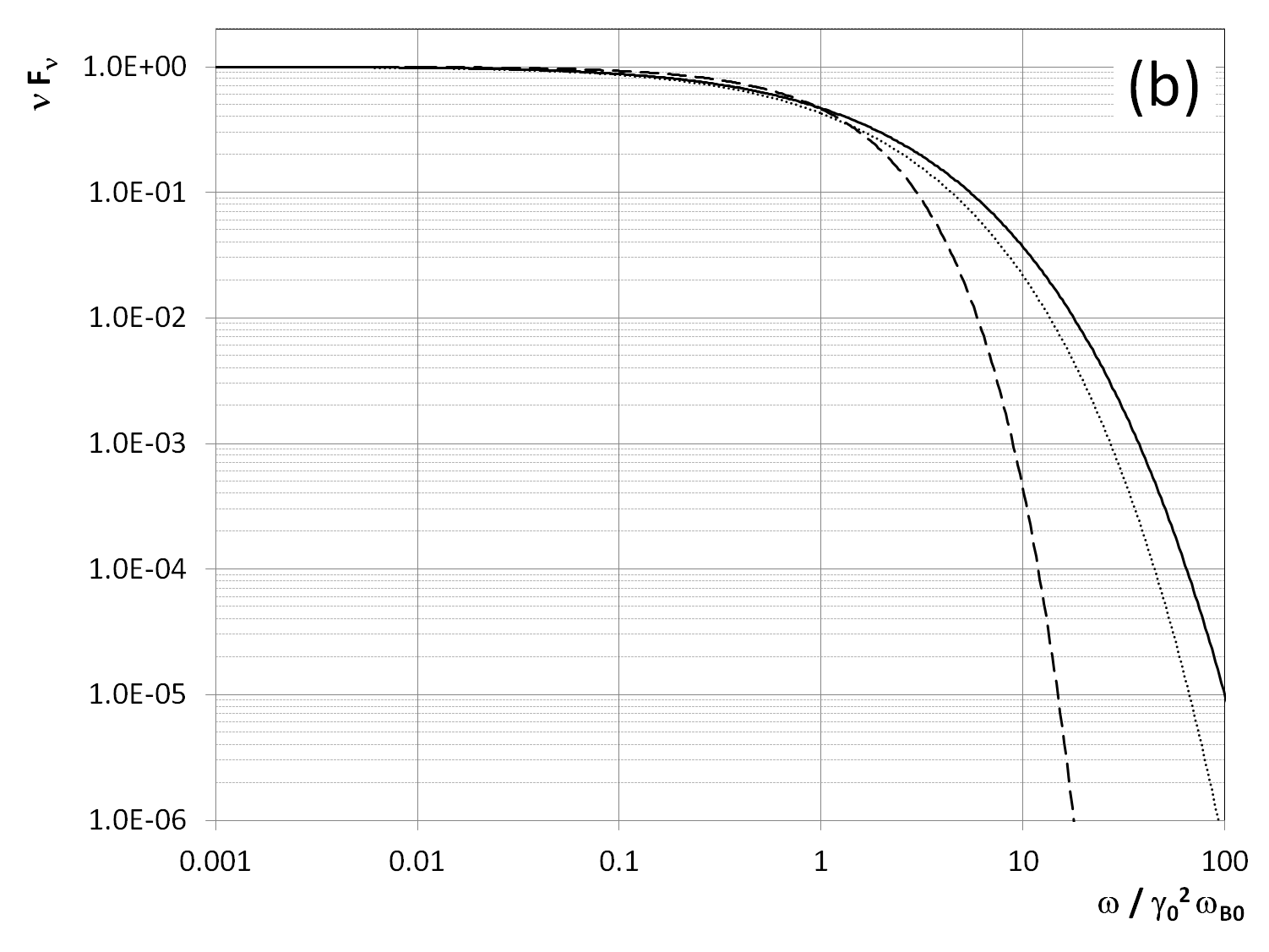}
    \caption{Spectral energy distributions ($\nu F_\nu$) of synchrotron radiation for power-law electron distribution with exponential cut-off ($\beta=1$, panel a) and with Gaussian cut-off ($\beta=2$, panel b) for the turbulent magnetic field (solid line) and for the uniform magnetic field in monochromatic approximation (dashed line) and calculated using approximate pitch-angle averaged emissivity function from \cite{approxAngleAvg} (dotted line). The power-law index of electron distribution is $p=3$. Vertical axis -- $\nu F_\nu$ in arbitrary units (logarithmic scale). Horizontal axis -- frequency (logarithmic scale).}
   \label{PL-Cut}
\end{figure}

In Fig.~(\ref{PL-Cut})  we compare synchrotron SEDs from power-law electron distribution with a high energy cut-off in the turbulent magnetic field to those in constant-strength magnetic field, obtained either in monochromatic approximation or using approximate pitch-angle averaged emissivity function from \cite{approxAngleAvg}. The comparison shows that the monochromatic approximation is not viable beyond the cut-off frequency.

The two expressions for synchrotron spectrum given by Eqs.~(\ref{standard_spectrum}) and (\ref{turbylentSy}), have distinct areas of application. If irregular component of the magnetic field is weaker than the regular one, then use of Eq.~(\ref{standard_spectrum}) is justified with appropriate integration over pitch angles (in nearly uniform magnetic field electrons' distribution over pitch angles may not necessarily be isotropic). If, on the contrary, the magnetic field is irregular, then we suggest using Eq.~(\ref{turbylentSy}) for the synchrotron spectrum. It is fully justified in the case of quasi-linear magnetic turbulence, but may still be better approximation in the case of strongly non-linear magnetic turbulence in addition to providing  simpler to handle expression.

\section{Summary}  \label{sec:summary}

In this paper we find -- in terms of elementary functions -- the exact expression 
for the spectrum of synchrotron radiation of an electron in turbulent magnetic field with Gaussian statistics of local magnetic field strengths. 

This expression reads
\begin{equation}
    \left( \frac{{\rm d} \dot{N}}{{\rm d} \omega} \right)_t
    =  \frac{\alpha}{3} \frac{1}{\gamma^2} \left( 1  + \frac{1}{x^{2/3}} \right)  \exp \left( - 2 x^{2/3} \right)  \, ,  
\end{equation}
where 
\begin{equation*}
x = \frac{\omega}{\omega_0}\, , \qquad  
\omega_0 = \frac{4}{3} \gamma^2 \frac{eB_0}{m_e c} \equiv \frac{4}{3} \gamma^2 \omega_\mathrm{_{B,0}}\, .
\end{equation*}

One should note the slower decline at high frequencies, $\propto \exp( -2x^{2/3})$  compared to the case of uniform magnetic field, where the decline is exponential, $\propto \exp( -8x/9)$. 
Turbulent magnetic field that satisfies Gaussian statistics is expected to occur naturally wherever there is quasilinear turbulence. 

Building on the simple expression for the spectrum of individual electron, we find -- in terms of elementary functions -- exact expression for the synchrotron spectrum from power-law electron distribution in the turbulent magnetic field given by Eq.~(\ref{uncutPL_exact}) and show that this spectrum can be reproduced using monochromatic approximation (each electron radiates at a single frequency proportional to the square of its Lorentz factor) with appropriate choice of the frequency (Eqs.~\ref{omega_m} and \ref{a_m}). We also derive the synchrotron spectrum for power-law electron distribution with cut-off (Eqs.~\ref{PL_spectrum} and \ref{Q_PL}). In the latter case we provide asymptotic expressions in terms of elementary functions both for low (Eq.~\ref{asymp_at_small_x}) and large (Eq.~\ref{asymp_at_large_x}) frequencies, as well as approximation valid for any frequency (Eq.~\ref{two_sided_approximation}). 

Again, Gaussian magnetic field strength fluctuations result in slower decline of the synchrotron spectrum beyond the cut-off, $\displaystyle \propto \exp\left( -a \left[ \frac{\omega}{\gamma_0^2 \omega_\mathrm{_{B,0}}} \right]^{2\beta/(3\beta+4)} \right)$, compared to $\displaystyle \propto \exp\left( - b \left[ \frac{\omega}{\gamma_0^2 \omega_\mathrm{_B}} \right]^{\beta/(\beta+2)} \right)$ in the case of constant-strength magnetic field, both assuming $\propto \exp\left( -\left[ \gamma/\gamma_0 \right]^\beta\right)$ type cut-off in the parent electron distribution\footnote{
Note different numerical coefficients in the exponential terms, $\displaystyle a = 2^{(\beta-2)/(\beta+2)} (\beta+2) (3\beta)^{-\beta/(\beta+2)}$ and $\displaystyle b = (3/4)^{2\beta/(3\beta+4)} \left(2+\frac{8}{3\beta} \right) \left(\frac{3\beta}{8} \right)^{4/(3\beta+4)}$. As $\beta$ changes from 1 to 2, the coefficients $a$ and $b$ vary between $\simeq 2.45$ and $\simeq 2.65$ and between $\simeq 1.63$ and $\simeq 1.67$, respectively; their ratio $a/b$ monotonically increases from $\simeq 1.49$ to $\simeq 1.62$.
}.

\acknowledgments

This research is supported by the Russian Foundation for Basic Research grant No 17-02-00525 (E.D.).



\appendix

\section{Derivation of Q(x)}
\label{sec:appendix1}

The simplest way to evaluate the integral in definition of $Q(x)$ is by splitting it into two integrals using recurrence relations for Macdonald functions. Both the integrals are given in \cite{IntegralsPrudnikov} (Eqs.~2.16.8.14 and 2.16.8.16) as simplified special cases of more general integral (Eq.~2.16.8.13). Unfortunately, both have typos in numerical coefficients. We therefore start with the 

general expression (Eq.~2.16.8.13 in \cite{IntegralsPrudnikov}) in the form 
\begin{multline} \label{GeneralIntegral}
    \int_0^{\infty} x^{\alpha-1} \exp\!\left(-\frac{p}{x^2}\right) K_{\nu}(cx) \, \mathrm{d}x =\\
    = \frac{2^{\alpha-2}}{c^{\alpha}}\, \Gamma\!\left(\frac{\alpha+\nu}{2}\right) \Gamma\!\left(\frac{\alpha-\nu}{2}\right)
        \, {}_0F_2\left(1-\frac{\alpha+\nu}{2},\, 1-\frac{\alpha-\nu}{2};\, -\frac{c^2 p}{4}\right) \\
    + \frac{c^{\nu} p^{(\alpha+\nu)/2}}{2^{\nu+2}}\, \Gamma\!\left(-\nu\right) \Gamma\!\left(-\frac{\alpha+\nu}{2}\right)
        \, {}_0F_2\left(1+\frac{\alpha+\nu}{2},\, 1+\nu;\, -\frac{c^2 p}{4}\right) \\
    + \frac{c^{-\nu} p^{(\alpha-\nu)/2}}{2^{2-\nu}}\, \Gamma\!\left(\nu\right) \Gamma\!\left(\frac{\nu-\alpha}{2}\right)
        \, {}_0F_2\left(1+\frac{\alpha-\nu}{2},\, 1-\nu;\, -\frac{c^2 p}{4}\right) \, .
\end{multline}
Here 
\begin{equation}
    {}_0F_2(a_1,a_2;z) = 
    \sum_{k=0}^{\infty}\frac{1}{{\left(a_{1}\right)_{k}}{\left(a_{2}\right)_{k}}}\frac{z^{k}}{k!}.
\end{equation}
is a generalized hypergeometric function and
\begin{equation}
\begin{array}{l}
    {\left(a\right)_{0}}=1\, , \\
    {\left(a\right)_{n}}=a(a+1)(a+2)\cdots(a+n-1)
\end{array}
\end{equation}
the Pochhammer’s symbol. In our case ($\alpha=1$, $c=1$, $\nu=5/3$) Eq.~\ref{GeneralIntegral} simplifies to become (we make substitution $p=4z$)
\begin{multline} \label{Integral}
    \int_0^{\infty} \exp\!\left(-\frac{4z}{\xi^2}\right) K_{5/3}(\xi) \, \mathrm{d}\xi =\\
    = \frac{1}{2}\, \Gamma\!\left(\frac{4}{3}\right) \Gamma\!\left(-\frac{1}{3}\right)
        \, {}_0F_2\left(-\frac{1}{3},\, \frac{4}{3};\, -z\right) \\
    + \frac{z^{4/3}}{2}\, \Gamma\!\left(-\frac{5}{3}\right) \Gamma\!\left(-\frac{4}{3}\right)
        \, {}_0F_2\left(\frac{7}{3},\, \frac{8}{3};\, -z\right) \\
    + \frac{z^{-1/3}}{2}\, \Gamma\!\left(\frac{5}{3}\right) \Gamma\!\left(\frac{1}{3}\right)
        \, {}_0F_2\left(\frac{2}{3},\, -\frac{2}{3};\, -z\right) \\
    = \frac{\pi}{\sqrt{3}}\,
        \left[ 
        -\, {}_0F_2\left(-\frac{1}{3},\, \frac{4}{3};\, -z\right) 
        + \frac{81}{40} z^{4/3} {}_0F_2\left(\frac{7}{3},\, \frac{8}{3};\, -z\right) \right.\\
        \left. + \frac{2}{3} z^{-1/3}  {}_0F_2\left(\frac{2}{3},\, -\frac{2}{3};\, -z\right) 
        \right]\, ,
\end{multline}
where we used the functional equation $\Gamma(z+1)=z\Gamma(z)$ to express all three products of gamma functions in terms of 
$\Gamma\!\left(\frac{1}{3}\right) \Gamma\!\left(\frac{2}{3}\right)$
and then Gauss’s multiplication formula $\prod_{k=1}^{n-1}\Gamma\left(\frac{k}{n}\right)=(2\pi)^{(n-1)/2}n^{-1/2}$.

Then we transform the three generalized hypergeometric series from Eq.~\ref{Integral}:
\begin{multline} \label{GGF1}
    {}_0F_2\left(-\frac{1}{3},\, \frac{4}{3};\, -z\right) = \\
    = 1 + \sum_{k=1}^{\infty}\frac{(-z)^{k}}{
    \left[\left(-\frac{1}{3}\right)\left(\frac{2}{3}\right)\cdots\left(-\frac{1}{3}+k-1\right)\right]\times
    \left[\left(\frac{4}{3}\right)\left(\frac{7}{3}\right)\cdots\left(\frac{4}{3}+k-1\right)\right]\times
    \left[1 \cdot 2 \cdots k\right]    }\\
    = 1 + \sum_{k=1}^{\infty}\frac{\left(\frac{1}{3}\right) \left(k-\frac{1}{3}\right) (-z)^{k}}{
    \left(-\frac{1}{3}\right) \left[\left(\frac{1}{3}\right)\left(\frac{2}{3}\right)\left(\frac{3}{3}\right)\cdots\left(k+\frac{1}{3}\right)\right]    }
    = 1 - \sum_{k=1}^{\infty}\frac{3^{3k} \left(3k-1\right) (-z)^{k}}{(3k+1)!}\\
    = - \sum_{k=0}^{\infty}\frac{\left(3k-1\right) }{(3k+1)!} \left(-3 z^{1/3}\right)^{3k} \, ,
\end{multline}
\begin{multline} \label{GGF2}
    {}_0F_2\left(\frac{7}{3},\, \frac{8}{3};\, -z\right) = \\
    = 1 + \sum_{k=1}^{\infty}\frac{(-z)^{k}}{
    \left[\left(\frac{7}{3}\right)\left(\frac{10}{3}\right)\cdots\left(\frac{7}{3}+k-1\right)\right]\times
    \left[\left(\frac{8}{3}\right)\left(\frac{11}{3}\right)\cdots\left(\frac{8}{3}+k-1\right)\right]\times
    \left[1 \cdot 2 \cdots k\right]    }\\
    = 1 + \sum_{k=1}^{\infty}\frac{\left(\frac{1}{3}\right) \left(\frac{4}{3}\right) \left(\frac{2}{3}\right) \left(\frac{5}{3}\right) (k+1) (-z)^{k}}{
    \left(\frac{1}{3}\right)\left(\frac{2}{3}\right)\left(\frac{3}{3}\right)\cdots\left(k+\frac{5}{3}\right)    }
    = 1 + 40 \sum_{k=1}^{\infty}\frac{3^{3k} (3k+3) (-z)^{k}}{(3k+5)!} \\
    = 40 \sum_{k=0}^{\infty}\frac{(3k+3)}{(3k+5)!} \left(-3 z^{1/3}\right)^{3k} 
    = 40 \sum_{k=0}^{\infty}\frac{3k}{(3k+2)!} \left(-3 z^{1/3}\right)^{3k-3} \, ,
\end{multline}
\begin{multline} \label{GGF3}
    {}_0F_2\left(\frac{2}{3},\, -\frac{2}{3};\, -z\right) = \\
    = 1 + \sum_{k=1}^{\infty}\frac{(-z)^{k}}{
    \left[\left(\frac{2}{3}\right)\left(\frac{5}{3}\right)\cdots\left(\frac{2}{3}+k-1\right)\right]\times
    \left[\left(-\frac{2}{3}\right)\left(\frac{1}{3}\right)\cdots\left(-\frac{2}{3}+k-1\right)\right]\times
    \left[1 \cdot 2 \cdots k\right]    }\\
    = 1 + \sum_{k=1}^{\infty}\frac{\left(k-\frac{2}{3}\right) (-z)^{k}}{
    \left(-\frac{2}{3}\right) \left[\left(\frac{1}{3}\right)\left(\frac{2}{3}\right)\left(\frac{3}{3}\right)\cdots\left(k\right)\right]    }
    = 1 - \frac{1}{2} \sum_{k=1}^{\infty}\frac{3^{3k} \left(3k-2\right) (-z)^{k}}{(3k)!}\\
    =  - \frac{1}{2} \sum_{k=0}^{\infty}\frac{\left(3k-2\right)}{(3k)!}  \left(-3 z^{1/3}\right)^{3k} \, .
\end{multline}
In Eq.\ref{GGF2} we formally added extra term to the series, which identically equals to 0 for any $z \neq 0$. 

After substituting Eqs. \ref{GGF1} -- \ref{GGF3} into Eq.~\ref{Integral} we note, that the three terms in square brackets represent different parts of a single series, then split this series into two (re-arranging terms), and eventually obtain
\begin{multline} \label{integral_compacted}
    \int_0^{\infty} \exp\!\left(-\frac{4z}{\xi^2}\right) K_{5/3}(\xi) \, \mathrm{d}\xi =\\
    = \frac{\pi}{\sqrt{3}}\,
        \left[ 
        \sum_{k=0}^{\infty}\frac{\left(3k-1\right) }{(3k+1)!} \left(-3 z^{1/3}\right)^{3k} 
        + \sum_{k=0}^{\infty}\frac{3k}{(3k+2)!} \left(-3 z^{1/3}\right)^{3k+1} \right.\\
        \left. + \sum_{k=0}^{\infty}\frac{\left(3k-2\right)}{(3k)!}  \left(-3 z^{1/3}\right)^{3k-1} 
        \right]\\
    = \frac{\pi}{3\sqrt{3}}\, z^{-1/3} \sum_{k=0}^{\infty}\frac{2-k}{k!}  \left(-3 z^{1/3}\right)^{k}
    = \frac{\pi}{3\sqrt{3}} \left( 2 z^{-1/3}  + 3 \right) \sum_{k=0}^{\infty}\frac{1}{k!}  \left(-3 z^{1/3}\right)^{k} \\ 
    = \frac{\pi}{\sqrt{3}} \left( \frac{2}{3 z^{1/3}}  + 1 \right) \exp\! \left(-3 z^{1/3}\right) . 
\end{multline}

The final result is derived from Eq.~\ref{integral_compacted} by substitution $z=8x^2/27$:
\begin{multline} 
    \frac{\sqrt{3}}{\pi} \int_0^{\infty} \exp\!\left(-\frac{32\, x^2}{27\, \xi^2}\right) K_{5/3}(\xi) \, \mathrm{d}\xi = \\
    = \left( \frac{2}{3 (8x^2/27)^{1/3}}  + 1 \right) \exp\! \left(-3 (8x^2/27)^{1/3}\right) \\ 
    = \left( \frac{1}{x^{2/3}}  + 1 \right) \exp\! \left(-2 x^{2/3}\right) . 
\end{multline}

\section{Laplace's method of estimating integrals}
\label{sec:appendix2}

Consider integral 
\begin{equation}
    I = \int_0^\infty g(x) \exp{\left(f(x)\right)}\, {\rm d}x \, ,
\end{equation}
such that the exponent has a sharp maximum at the point $x_0$, where function $f(x)$ reaches it maximum. Approximate integration (by Laplace's method) can be done by replacing $f(x)$ by two leading terms of its Taylor series expansion in the vicinity of $x=x_0$:
\begin{equation}
    f(x) \simeq f(x_0) + \frac{1}{2} f^{\prime \prime}(x_0)\, (x-x_0)^2\, .
\end{equation}
The exponent then becomes a Gaussian function, so that
\begin{multline} \label{LaplaceIntegration}
    I \simeq \exp{\left(f(x_0)\right)} 
    \int_0^\infty g(x) \exp{\left(-\frac{(x-x_0)^2}{2/\lvert f^{\prime \prime}(x_0) \rvert}\right)}\, {\rm d}x \\
    \simeq \exp{\left(f(x_0)\right)}\, g(x_0) \left( \frac{2\pi}{\lvert f^{\prime \prime}(x_0) \rvert}\right)^{1/2} \, .
\end{multline}

\newpage

\bibliography{references}{}
\bibliographystyle{aasjournal}

\end{document}